\documentclass[reprint,aps,prb,amsmath,superscriptaddress,secnumarabic,amssymb,nobibnotes,longbibliography]{revtex4-1}
\usepackage{graphicx}

\begin{document}

\title{Integrating all-optical switching with spintronics}

\author{M.L.M. Lalieu}
\email[Corresponding author: ]{m.l.m.lalieu@tue.nl}
\affiliation{Department of Applied Physics, Institute for Photonic Integration, Eindhoven University of Technology, P.O. Box 513, 5600 MB Eindhoven, The Netherlands}
\author{R. Lavrijsen}
\affiliation{Department of Applied Physics, Institute for Photonic Integration, Eindhoven University of Technology, P.O. Box 513, 5600 MB Eindhoven, The Netherlands}
\author{B. Koopmans}
\affiliation{Department of Applied Physics, Institute for Photonic Integration, Eindhoven University of Technology, P.O. Box 513, 5600 MB Eindhoven, The Netherlands}

\begin{abstract}
All-optical switching (AOS) of magnetic materials describes the reversal of the magnetization using short (femtosecond) laser pulses, and has been observed in a variety of materials. In the past decade it received extensive attention due to its high potential for fast and energy-efficient data writing in future spintronic memory applications. Unfortunately, the AOS mechanism in the ferromagnetic multilayers commonly used in spintronics needs multiple pulses for the magnetization reversal, losing its speed and energy efficiency. Here, we experimentally demonstrate `on-the-fly' single-pulse AOS in combination with spin Hall effect (SHE) driven motion of magnetic domains in Pt/Co/Gd synthetic-ferrimagnetic racetracks. Moreover, using field-driven-SHE-assisted domain wall (DW) motion measurements, both the SHE efficiency in the racetrack is determined and the chirality of the optically written DW's is verified. Our experiments demonstrate that Pt/Co/Gd racetracks facilitate both single-pulse AOS as well as efficient SHE induced domain wall motion, which might ultimately pave the way towards integrated photonic memory devices.
\end{abstract}
\maketitle

The high potential of all-optical switching for fast and energy-efficient memory devices was quickly recognized after it was first discovered in GdFeCo alloys about a decade ago \cite{Stanciu2007}. With spintronic integration in prospect, the discovery initiated a rapidly developing field of research, initially aimed at unraveling the mechanism of this ultrafast switch. Soon it was discovered that the AOS in rare earth-transition metal (RE-TM) alloys is a purely thermal single-pulse process \cite{Ostler2012,Radu2011}, and that an earlier observed helicity dependence was the result of magnetic circular dichroism \cite{Stanciu2007,Khorsand2012}. The research field gained an additional boost when AOS was observed in ferromagnetic thin films and multilayers \cite{Lambert2014}, which are material systems already heavily used in the field of spintronics for future memory devices such as the racetrack memory \cite{Parkin2008,Ryu2012,Yang2015} and next generation magnetic random access memory \cite{Fukami2016}. Unfortunately, the helicity-dependent AOS found in these materials turned out to be a cumulative process needing multiple pulses \cite{Hadri2016,Medapalli2017}, preventing its use in fast spintronic devices. Clearly, the thermal single-pulse AOS mechanism is needed for successful spintronic integration. Although this mechanism is well established in RE-TM alloys, spintronic devices like the racetrack memory rely on interface-induced phenomena inherent to multilayered ultra-thin-film structures. In our recent work, we have shown efficient single-pulse AOS in such multilayers, made of a Pt/Co/Gd synthetic-ferrimagnetic stack \cite{Lalieu2017-2}. This structure was chosen because of; (i) the interfacial anti-ferromagnetic coupling between the Co and Gd layers \cite{Pham2016}, (ii) the large contrast in demagnetization times between the Co and Gd \cite{Koopmans2010,Wietstruk2011}, (iii) the Pt seed layer induced perpendicular magnetic anisotropy, and (iiii) the built-in interfacial Dzyaloshinskii-Moriya interaction (iDMI)\cite{Pham2016}. 

In this Letter, we experimentally demonstrate that the Pt/Co/Gd stack is indeed an ideal candidate to facilitate the integration of AOS with spintronics, and more specifically with the racetrack memory. We do this by demonstrating clear and robust single-pulse AOS in Pt/Co/Gd racetracks, and verifying that the DW's of the optically written domain are chiral Ne\'el walls that can be moved coherently along the racetrack by means of the SHE. Moreover, the SHE efficiency in the Pt/Co/Gd racetrack is determined, predicting high DW velocities, and a proof-of-concept measurement is presented demonstrating on-the-fly data writing in the racetrack, i.e. showing single-pulse AOS while simultaneously sending an electrical current through the racetrack that transports the written magnetic domains by means of the SHE.

The measurements were performed on Ta(4) / Pt(4) / Co(1) / Gd(3) / Pt(2) stacks (thickness in nm), which were deposited on a Si/SiO$_{2}$(100 nm) substrate using DC magnetron sputtering (see Methods). The samples were patterned into $5 \ \mu$m wide and $90 \ \mu$m long wires using electron beam lithography and argon ion milling. Each wire contains typically two sets of lateral legs ($2 \ \mu$m wide) forming a Hall cross, used to measure the out-of-plane magnetization by means of the anomalous Hall effect (AHE). All structures showed perpendicular magnetic anisotropy with square out-of-plane hysteresis loops and 100\% remanence.

\begin{figure}
	\includegraphics[scale=0.28]{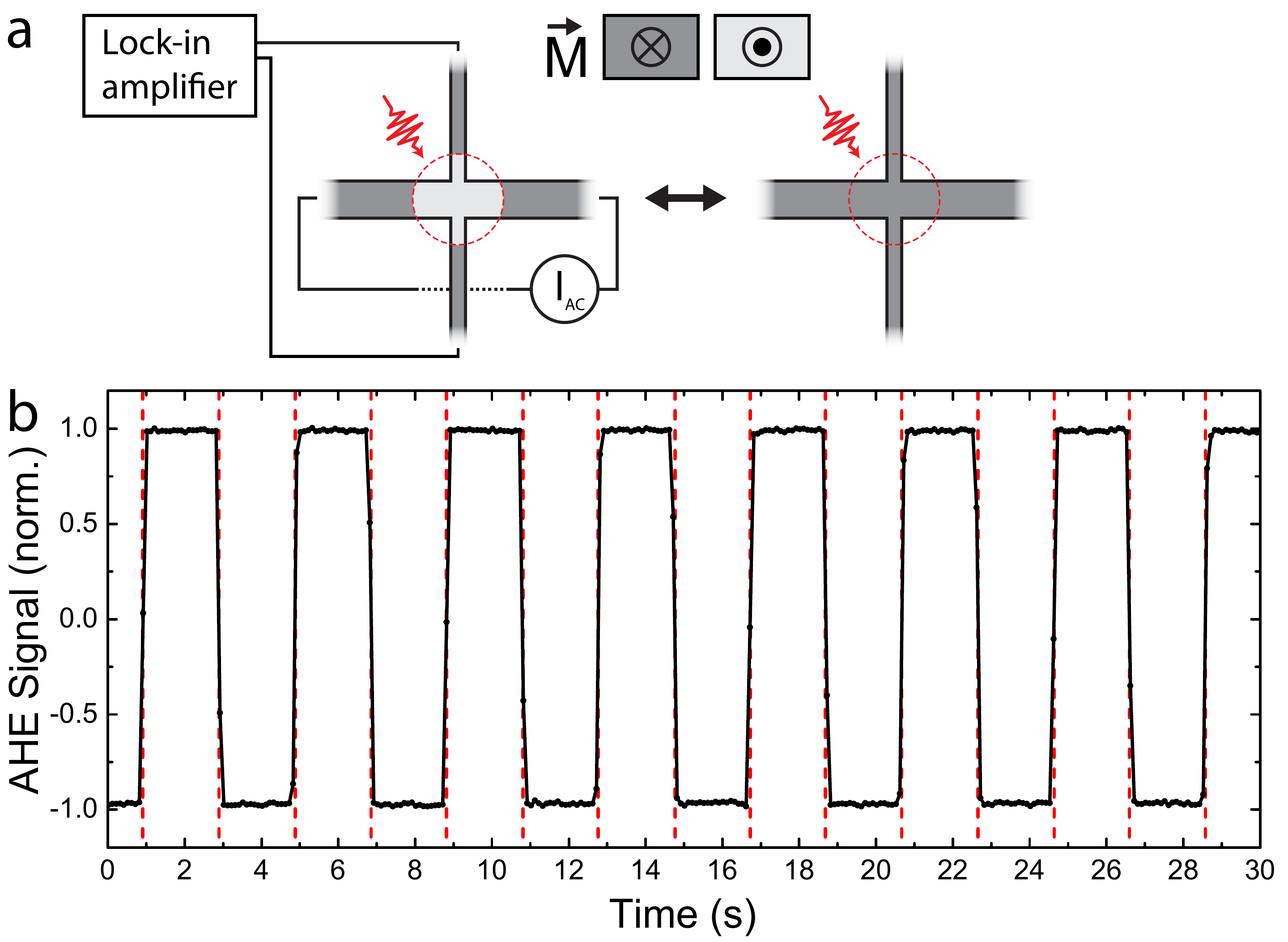}
	\caption{\textbf{Deterministic single-pulse AOS in a Pt/Co/Gd racetrack} (a) Schematic overview of the AOS measurement in a Hall cross on the Pt/Co/Gd wire. A small AC current is applied along the wire, while the resulting anomalous Hall voltage is measured across the legs using a lock-in amplifier. Exciting the cross by subsequent single linearly polarized laser pulses toggles the magnetization in the exposed region (dotted circle) up and down. (b) Measurement of the normalized AHE signal as a function of time during laser-pulse excitation at a repetition rate of $0.5$ Hz. No external field was applied during the measurement.}
	\label{Fig:ToggleAOSHallCross}
\end{figure}

First, the AOS in the magnetic wires was investigated by measuring the magnetization in one of the Hall crosses while at the same time it was exposed to a train of linearly polarized laser pulses ($\approx 100$ fs). The magnetization in the cross was measured using the AHE (see Methods), as illustrated in Fig.\ \ref{Fig:ToggleAOSHallCross}(a). The AHE signal is proportional to the out-of-plane component of the magnetization in the Hall cross area, which was normalized to the up ($+1$) and down ($-1$) saturation values using an external out-of-plane magnetic field. A typical measurement (without any external field) is shown in Fig.\ \ref{Fig:ToggleAOSHallCross}(b). In order to clearly identify the single pulses in the AHE measurement a relatively low laser-pulse repetition rate of $0.5$ Hz was used. It is clearly seen that the magnetization in the Hall cross region toggles between the saturated up ($+1$) and down ($-1$) states at the frequency of the incoming laser pulses. Repeating the measurement for a longer time demonstrated a $100\%$ success rate of the AOS for over more than $5000$ subsequent laser pulses.  This shows that indeed a deterministic single-pulse all-optical switch of the magnetization is present in the patterned Pt/Co/Gd wires.

The result presented in Fig.\ \ref{Fig:ToggleAOSHallCross}(b) shows full AOS in the Hall cross. The laser spot was larger than the Hall cross, meaning that the area of the Pt/Co/Gd wire that was exposed to the laser pulse was larger than the region probed by the Hall cross. It is known that depending on the laser fluence, a multidomain state can form at the centre of the (Gaussian shaped) laser spot, in which case only AOS is observed in an outer rim of the excited area \cite{Hadri2016}. In view of the transport measurements discussed in the following, it is important that a single homogeneous domain is written in the wire by the laser pulse. Supplementary Note 1 shows measurements on the AOS as a function of the overlap of the laser spot and the Hall cross, demonstrating that indeed homogeneous domains were written in the Pt/Co/Gd wire.

With the AOS in the wires verified, transport measurements on the optically written domains were performed. The DW's in the Pt/Co/Gd wire are expected to be chiral Ne\'el walls due to the iDMI \cite{Pham2016}. Such chiral Ne\'el walls can be moved coherently through the wire using an electrical current, exploiting the SHE in the heavy-metal Pt seed layer of the Pt/Co/Gd wire. The SHE in the Pt layer results in a spin accumulation at the Pt/Co interface that generates a torque on the DW that causes it to move. The direction of the DW motion is determined by the sign of the SHE and the chirality of the DW. For a ferromagnet on top of a Pt layer, the SHE driven DW motion is reported to be along the current direction, i.e. against the electron flow direction\cite{Emori2013,Ryu2013}.

\begin{figure}
	\includegraphics[scale=0.35]{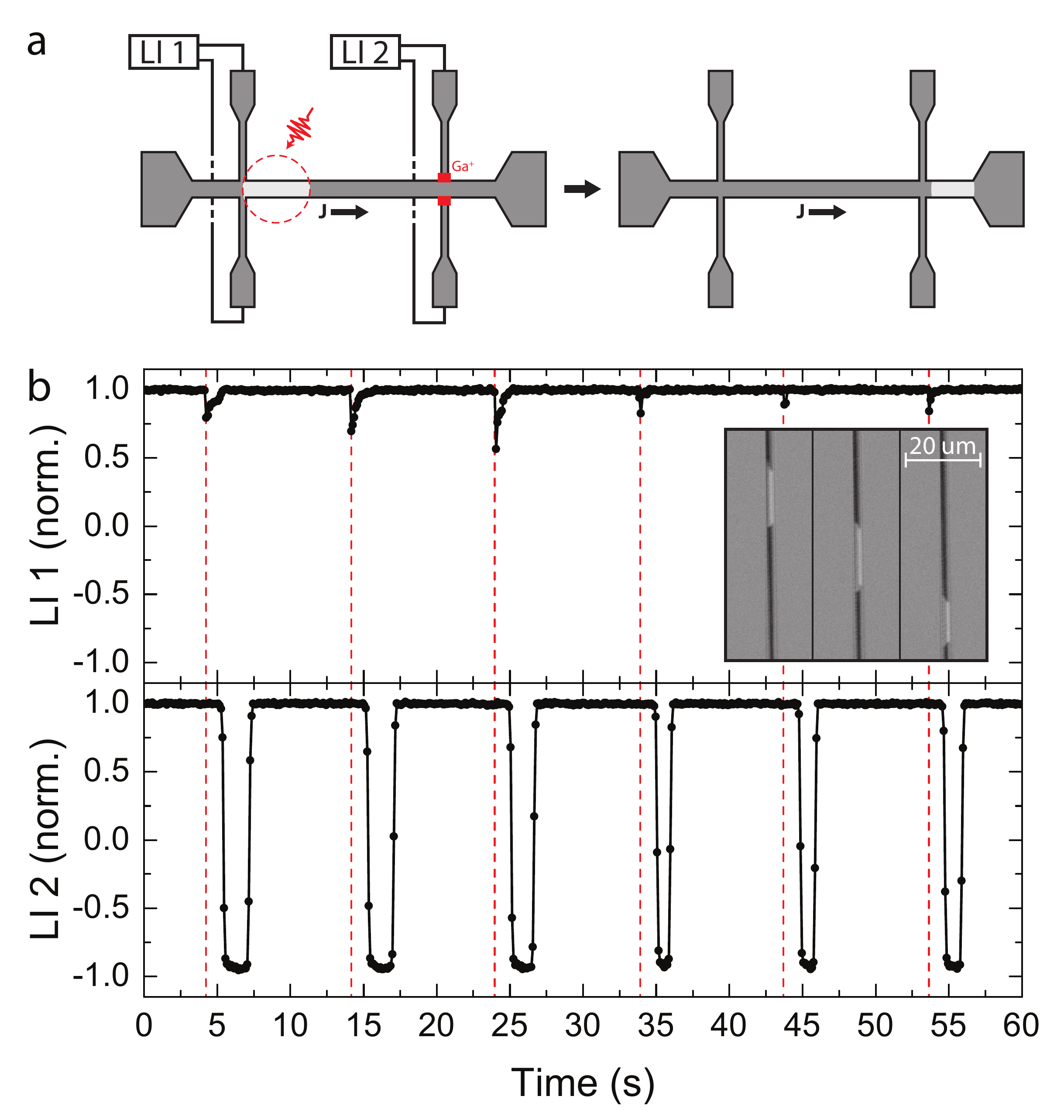}
	\caption{\textbf{On-the-fly AOS with purely SHE-driven DW motion} (a) Illustration of the field free proof-of-concept measurement, demonstrating on-the-fly single-pulse AOS in combination with purely SHE-driven transport of the optically written domains in a Pt/Co/Gd wire. The AHE signal in the Hall crosses is measured using lock-in amplifiers LI $1$ and LI $2$. The red dotted circle illustrates the region exposed by the laser pulse, and the red blocks indicate the regions exposed to Ga$^{+}$ ion irradiation. (b) Results of the measurement illustrated in (a), showing the normalized AHE signal of the two Hall crosses as a function of time. The red dotted lines mark the time at which the laser pulse hits the wire and writes the magnetic domain.}
	\label{Fig:ProofOfPrinciple}
\end{figure}

\begin{figure*}
	\includegraphics[scale=0.32]{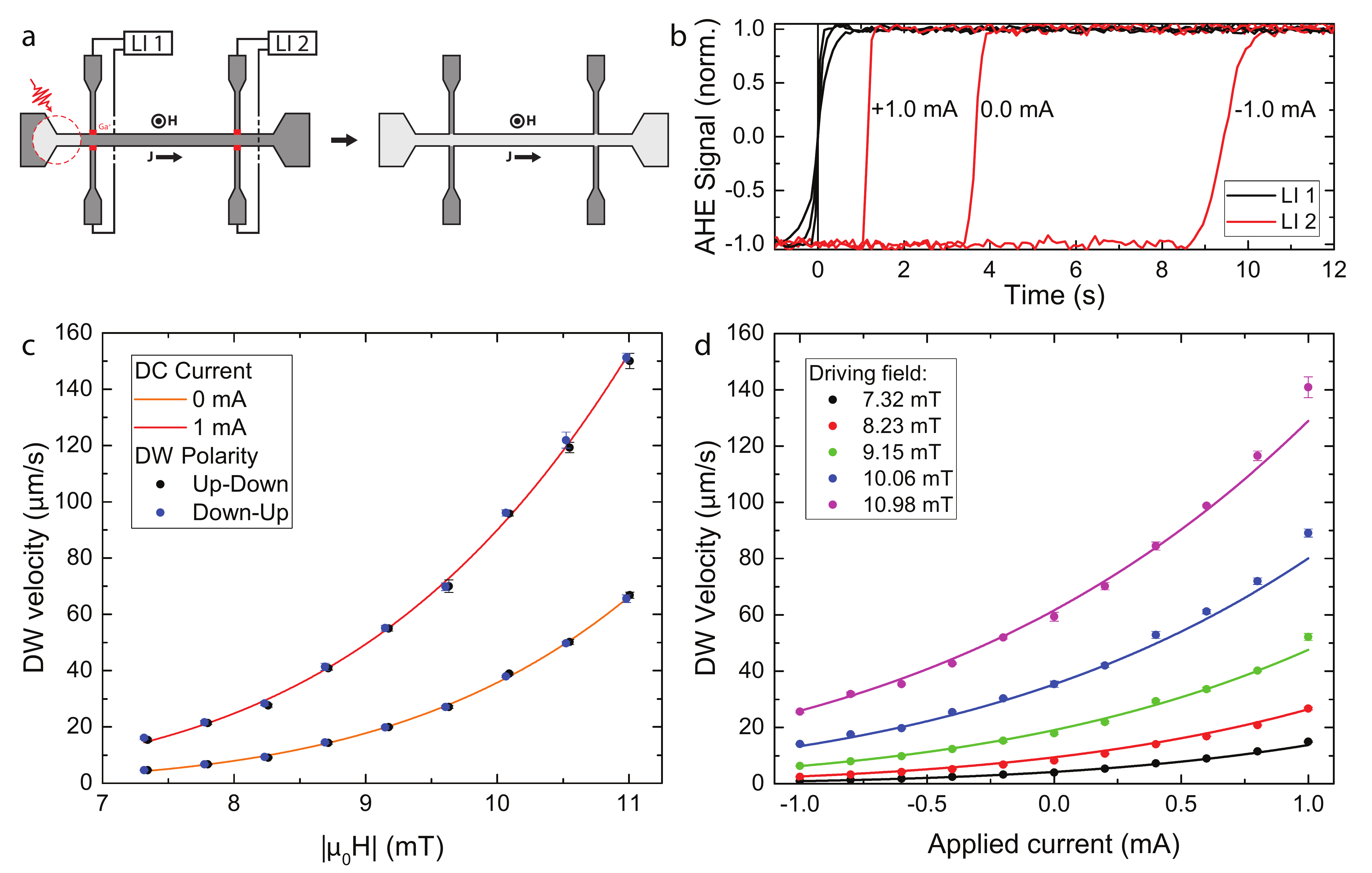}
	\caption{\textbf{Field-driven-SHE-assisted DW velocity measurements} (a) Schematic overview of the field-driven-SHE-assisted DW motion measurement on the Pt/Co/Gd wire. The AHE signal in the Hall crosses is measured using lock-in amplifiers LI $1$ and LI $2$. The red dotted circle illustrates the region exposed by the laser pulse, and the red blocks indicate the regions exposed to Ga$^{+}$ ion irradiation. (b) Typical time of flight measurements for three different applied current amplitudes, showing the normalized AHE signal in both Hall crosses as a function of time. The step in the signal from $-1$ (down) to $+1$ (up) indicates that the DW is passing the specific Hall cross. (c) DW velocity for both DW polarities as a function of the driving field amplitude with $0$ and $+1$ mA of DC current sent through the wire (direction indicated in (a)). The solid lines represents fits to the data following the creep law (Eq.\ (\ref{Eq:Creep})). (d) DW velocity as a function of the current sent through the wire for different driving fields. Solid lines represent fits to the data using Eq.\ (\ref{Eq:Creep}).}
	\label{Fig:FDSHEADWM}
\end{figure*}

Combining the SHE-driven transport of the optically written domains with the single-pulse AOS in the racetrack, we have been able to establish on-the-fly data writing. In such measurement, illustrated in Fig.\ \ref{Fig:ProofOfPrinciple}(a), AOS is used to write a domain in a Pt/Co/Gd wire containing two Hall crosses, while at the same time a chosen DC current is flowing through the wire (direction indicated in figure). Since both DW's enclosing the written domain have the same chirality (shown in Fig.\ \ref{Fig:FDSHEADWM}(c)), they will move coherently along the current direction by the SHE as soon as they are written. This means that the full domain will be transported along the wire, passing the Hall cross at the end where it will be recorded using the AHE.

As indicated in Fig.\ \ref{Fig:ProofOfPrinciple}(a) (red squares), the legs of the second (right) Hall cross in the Pt/Co/Gd wire were exposed to Ga$^{+}$ ion irradiation\cite{Franken2012}. This was done to magnetically `cut-off' the legs in order to prevent DW pinning at the entrance of the cross, as is discussed in Supplementary Note 2. To avoid pinning of one of the DW's at the first Hall cross, the laser was aligned not at the centre but at the right side of the first Hall cross. In order to verify if and when a domain is written, a small overlap between laser pulse and the first Hall cross was maintained. Similar measurements performed with the laser aligned completely in between the two Hall crosses and with the laser centred at the centre of the first Hall cross can be found in Supplementary Note 3, of which the latter clearly demonstrates the pinning of the DW at the entrance of the non-irradiated Hall cross.

The result of the measurement with a laser-pulse repetition rate of $0.1$ Hz and a DC current of $+5.5$ mA is shown in Fig.\ \ref{Fig:ProofOfPrinciple}(b). In the top half of the figure, the normalized AHE signal of the first Hall cross is shown as a function of time. It can be seen that small peaks appear in the signal at the frequency of the laser pulses. These peaks start with a sudden step, corresponding to the (small) exposed part of the Hall cross being switched by the laser, whereafter the signal quickly returns to the saturation value, showing the DW moving out of the cross. More interesting is the AHE signal of the second Hall cross, shown in the bottom half of the figure. It can be seen that shortly after the domain is written (red dotted lines), the magnetization in the second Hall cross switches down ($-1$) and shortly thereafter switches back up ($+1$) again. These switches mark the passing DW's, i.e. the passing domain. The time between the two switches, which is a measure of the width of the domain, varies, which is attributed to random pinning of the DW's in the wire. This proof-of-concept measurement demonstrates on-the-fly single-pulse AOS and simultaneous SHE-driven motion of magnetic domains in a single racetrack. 

Additionally, the SHE induced DW motion was verified using a Kerr microscope, shown in the inset of Fig.\ \ref{Fig:ProofOfPrinciple}(b). The figure shows three snap shots of a magnetic up (white) domain in an otherwise down (black) magnetized Pt/Co/Gd wire ($2$ $\mu$m wide), while it was moved through the wire by a train of current pulses. As can be seen, the DW's indeed move coherently through the wire (against the electron flow), while a change in the domain width is observed as was discussed in the previous measurement.

A quantitative analysis of the SHE efficiency and DW chirality in the optically written domains was obtained by performing field-driven-SHE-assisted DW velocity measurements, as illustrated in Fig.\ \ref{Fig:FDSHEADWM}(a). In such measurement the DW motion is driven by an out-of-plane applied field while at the same time a DC current is sent through the wire. Depending on the current polarity, the DW motion is either assisted or hindered via the SHE, resulting in an increase or decrease of the DW velocity, respectively. The Pt/Co/Gd wires used for these measurements contain two Hall crosses that are separated by $60 \ \mu$m, and the legs of both crosses are exposed to Ga$^{+}$ ion irradiation (discussed earlier). At the start of the measurement the magnetization in the wire is saturated by the external field, whereafter a (static) field is applied in the opposite direction with an amplitude below the domain nucleation field, but above the DW propagation field. Using a single laser pulse, a domain is written into the wire left of the first Hall cross, see Fig.\ \ref{Fig:FDSHEADWM}(a). The applied field will cause the domain to expand through the wire, causing it to pass the two Hall crosses, which will be recorded by a switch in the AHE signal. Using the time of flight and the distance between the two crosses the DW velocity is determined. A typical measurement of the AHE signal in both Hall crosses as a function of time for three different DC current values is shown in Fig.\ \ref{Fig:FDSHEADWM}(b), where the passing of a DW through the Hall crosses is clearly seen in the AHE signal, and the effect of the SHE on the time of flight is apparent.

The DW chirality of the optically written domains was investigated by measuring the effect of the SHE on the DW velocity for both up-down and down-up DW polarities, which were obtained by reversing the saturation and propagation fields. Figure \ref{Fig:FDSHEADWM}(c) shows the DW velocity for both DW polarities as a function of the driving field amplitude with $0$ and $+1$ mA of DC current sent through the wire (direction indicated in Fig.\ \ref{Fig:FDSHEADWM}(a)). The solid lines represent a fit using the creep law for DW motion, which will be discussed later. More important are the observations that the DW velocity is independent of the DW polarity for both current amplitudes, and that a current of $+1$ mA increases the DW velocity with respect to the case without current. The latter affirms that the current induced contribution to the DW motion is against the direction of electron flow, as was also shown in Fig.\ \ref{Fig:ProofOfPrinciple}(b), indicating SHE driven DW motion dominated by the bottom Pt/Co interface as discussed earlier. The fact that both DW polarities are moved in the same direction at the same velocity by the SHE confirms that they are chiral Ne\'el walls as discussed earlier, indicating the presence of the iDMI in these wires.

Similar DW velocity measurements, now as a function of the current sent through the wire and for different driving fields $H_{\mathrm{ext}}$, were used to determine the SHE efficiency.  The results of these measurements are presented in Fig.\ \ref{Fig:FDSHEADWM}(d). The DW motion for the used driving fields of a few millitesla  is in the creep regime. In this regime the DW velocity can be described by \cite{Lee2011}
\begin{equation}
v = v_{0} \exp\left[-\chi\left(\mu_{0}H_{\mathrm{ext}} \pm \epsilon_{\mathrm{SHE}} J_{\mathrm{DC}}\right)^{-1/4}\right],
\label{Eq:Creep}
\end{equation}
in which $\epsilon_{\mathrm{SHE}}$ represents the SHE efficiency, defined as the current density to effective (out-of-plane) field conversion factor, $J_{\mathrm{DC}}$ the current density, $v_{0}$ the characteristic velocity and $\chi$ a scaling factor including the pinning potential and thermal energy. By doing field-driven DW motion measurements, the used current densities could be kept low enough to prevent a significant change in temperature by Joule heating, which was verified by four-point resistance measurements (not shown). As a result, the values of $v_{0}$ and $\chi$ can be assumed to be independent of the current density, and should be constant in the measurements presented in Fig.\ \ref{Fig:FDSHEADWM}(d). Therefore, the DW velocity as a function of the current density is fitted using Eq.\ (\ref{Eq:Creep}), where $H_{\mathrm{ext}}$ is known for each curve, and $v_{0}$, $\chi$ and $\epsilon_{\mathrm{SHE}}$ are used as global fit parameters. The fitted value for the SHE efficiency is equal to $\epsilon_{\mathrm{SHE}} = 9.73 \pm 0.08$ mT/($10^{11}$A/m$^{2}$) (using a homogeneous current distribution throughout the stack in the current density calculation). This value agrees well with the values found in literature for similar structures using different measurement methods \cite{Emori2012,Emori2013,Pai2016}.

Using the measured SHE efficiency a simple prediction can be made for the DW velocity that can be reached when driven by intense nanosecond (ns) current pulses as used in Ref. \cite{Yang2015}. In their work, the authors used current pulses with a current density up to $30 \cdot 10^{11}$ A/m$^{2}$ and a pulse duration of $5$ ns in similar sized wires. With the SHE efficiency of $\approx9.7$ mT/($10^{11}$A/m$^{2}$) in the present Pt/Co/Gd wires, this would result in an effective out-of-plane field of $\approx 290$ mT. This field can be related to a DW velocity using the work presented in Ref. \cite{Pham2016}, where the field-driven DW velocity is measured in an identical Pt/Co/Gd stack using out-of-plane field pulses with a duration of $20$ ns and strengths up to $300$ mT. Based on that work, a DW velocity as high as $700$ m/s can be extrapolated for the earlier mentioned current pulses. Moreover, it was recently shown that this DW velocity is expected to increase significantly when using a system with a more compensated magnetization \cite{Yang2015} or angular momentum \cite{Kim2017}, e.g. by reducing the Co thickness \cite{Lalieu2017-2}.

In conclusion, we have experimentally demonstrated that both single-pulse AOS as well as SHE induced domain wall motion can be combined in a Pt/Co/Gd racetrack with perpendicular magnetic anisotropy, exploiting the chiral Ne\'el structure of the DW's for coherent and efficient motion of the optically written domains. Ultimately, these results might pave the way towards integrated photonic memory devices, in which all-optical control of magnetism and spintronics meet.

\acknowledgments{This work is part of the Gravitation program 'Research Centre for Integrated Nanophotonics', which is financed by the Netherlands Organisation for Scientific Research (NWO).

\section*{Methods}
\subsection*{Sample fabrication}
The samples used in this work were deposited on Si substrates coated with $100$ nm of SiO$_{2}$. The deposition was done using DC magnetron sputtering at room temperature. The base pressure in the deposition chamber was $10^{-9}$ mbar. After deposition, Ti/Au contacts were deposited on top of the full sheet sample by a lift-off procedure using UV lithography (ma-N 415 photoresist). Lastly, $5 \ \mu$m wide and $90 \ \mu$m long wires were created using electron beam lithography to create a hard mask (ma-N 2410 photoresist) in combination with argon ion milling.

\subsection*{Measurement techniques}
The AOS was achieved using linearly polarized laser pulses with a central wavelength of $700$ nm and a pulse duration of $\approx 100$ fs. The pulse energy used throughout the work presented in the Letter was $\approx 12$ nJ, while the laser spot had a radius ($1/\mathrm{e}$ Gaussian pulse) of $\approx 20$ $\mu$m.

The out-of-plane magnetization in the Hall crosses was measured using the AHE. To measure the AHE signal a small AC current (150 $\mu$A) is sent through the wire, and the resulting anomalous Hall voltage across the lateral legs is measured using a lock-in amplifier. 

In case of the Kerr microscope images a differential technique was used. In this technique first an image of the magnetic wire with the magnetization saturated in either the up or down direction was captured. This (background) image was then subtracted from the subsequent images in order to enhance the magnetic contrast.

\subsection*{\large Supplementary Note 1: AOS as a function of the overlap between laser spot and Hall cross}

The result presented in Fig.\ 1(b) of the main Letter shows full AOS in the Hall cross. This was measured with the centre of the laser spot aligned to the centre of the Hall cross. As was mentioned, the laser spot was larger than the cross, meaning that the area of the Pt/Co/Gd wire that is exposed (and switched) by the laser pulse is larger than the region probed by the Hall cross. It is known that depending on the laser fluence, a multidomain state can form at the centre of the (Gaussian shaped) laser spot, in which case only AOS is observed in an outer rim of the excited area \cite{Hadri2016}. In this section, it is verified that a single homogeneous domain was written in the Pt/Co/Gd wire by the laser pulse.

In order to check that a homogeneous domain is written by the laser pulse, the measurement performed in Fig.\ 1 of the main Letter was repeated for different alignments of the laser spot with respect to the Hall cross. At each new alignment, the magnetization in the wire was first saturated using an externally applied field, whereafter the field was turned off and the Hall cross was exposed to a single laser pulse. When there is (partial) overlap between the Hall cross and the center area of the laser spot where the fluence $F\left(x,y\right)$ is above the AOS threshold fluence $F_{0}$, the magnetization in the Hall cross will be switched, which is recorded by a step in the AHE signal (similar as shown in Fig.\ 1(b) of the main Letter). The size of the AHE step is proportional to the area of the Hall cross that is switched by the laser. 

Figure\ \ref{Fig:AOSLineScan} shows the normalized AHE step size as a function of $x$ and $y$ position, where $\left(x,y\right) = \left(0,0\right)$ corresponds to the center of the (Gaussian) laser spot being aligned to the center of the Hall cross. The $x$ (black) and $y$ (red) scans are performed with $y = 0$ and $x = 0$, respectively, and each data point is an average of 7-8 subsequent measurements. When the laser spot is sufficiently far away from the Hall cross, i.e., for $\left|x\right|,\left|y\right|>15$ $\mu$m, the AHE step size is zero, meaning that there is no overlap between the Hall cross and the laser spot (or at least no sufficient overlap). Moving the laser spot closer to the Hall cross, i.e., decreasing $\left|x\right|$ or $\left|y\right|$, the AHE step size increases towards saturation at $\left|x\right|,\left|y\right|\approx5$ $\mu$m. The increase in AHE step size corresponds to the center part of the laser spot, where $F\left(x,y\right) \geq F_{0}$, moving into the Hall cross area. This is illustrated by the left and right cartoons in the figure in which the laser spot area with $F\left(x,y\right) \geq F_{0}$ (red dotted circle) overlaps $\approx30\%$ of the Hall cross area. For $\left|x\right|,\left|y\right|<5$ $\mu$m the AHE step size is constant and equal to saturation. The presence of these plateaus in both $x$ and $y$ scans demonstrates that for the full area of the laser spot where $F\left(x,y\right) \geq F_{0}$ there is full AOS. In other words, this means that indeed a homogeneous domain was written in the Pt/Co/Gd wire by the laser pulse. As a side note, the full width at half maximum of the curves are equal to the size of the written magnetic domain along the $x$ and $y$ directions, showing a domain size of $\approx 20$ $\mu$m and a slightly elliptically shaped laser spot, which was verified using wide field Kerr microscopy on a full-sheet sample. 

\begin{figure}
	\includegraphics[scale=0.35]{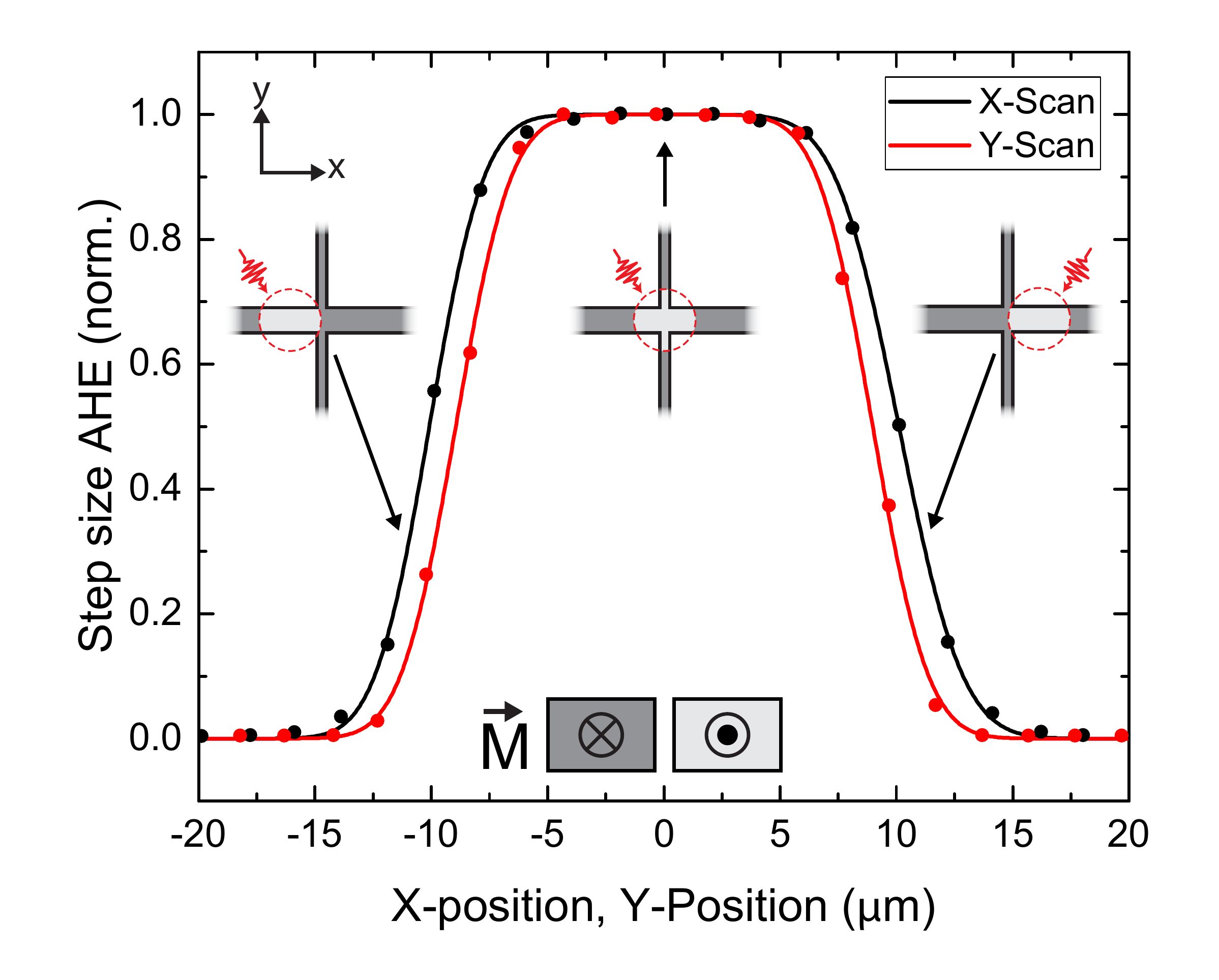}
	\caption{\textbf{AOS as a function of the overlap between laser spot and Hall cross.} Normalized step size of the AHE signal as a function of the $x$ and $y$ position. The step size in the AHE signal measured across the Hall cross is a measure of the area of the Hall cross that is switched by the laser pulse. $\left(x,y\right) = \left(0,0\right)$ corresponds to the center of the laser spot being aligned to the center of the Hall cross.}
	\label{Fig:AOSLineScan}
\end{figure}

Lastly, it is known that when the pulse energy is sufficiently increased, the fluence at the center of the laser spot can be increased to a value above a second threshold fluence. At this fluence the lattice temperature is heated above the Curie temperature, resulting in the formation of a multidomain state on cool down. Such a multidomain state at the center of the written domain would show up as a dip in the AHE step size around $\left(x,y\right) = \left(0,0\right)$ in the measurement presented in Fig.\ \ref{Fig:AOSLineScan}. Such a dip was indeed observed when repeating the measurement of Fig.\ \ref{Fig:AOSLineScan} with increasing pulse energy (not shown), verifying that inhomogineties in the written domain can indeed be measured, and thus confirming that for the laser pulses used in Fig.\ \ref{Fig:AOSLineScan} and the main Letter a homogeneous domain was written. 

\subsection*{\large Supplementary Note 2: DW pinning and Ga$^{+}$ irradiation}

In the main Letter it was mentioned that the legs of some of the Hall crosses were irradiated with Ga$^{+}$ ions in order to prevent pinning at the entrance of the cross. The pinning of the DW at the entrance of a non-irradiated Hall cross in a typical on-the-fly AOS measurement as performed in the main Letter is demonstrated in the measurement discussed in Supplementary Note 3 (Fig.\ \ref{Fig:ProofOfPrincipleOnCross}(c)). A visual presentation using a Kerr microscope is presented in the top row of Fig.\ \ref{Fig:GaIrradiation}. In this figure a down domain (dark) in an otherwise up (light) magnetized Pt/Co/Gd wire is located between two Hall crosses. Using three current pulses of alternating direction (see figure) it can be seen that the domain tries to move along the current direction, but gets pinned at the entrance of the cross it is moving towards. Moreover, it can be seen that the end points of the DW get pinned at the start of the legs, while the center of the DW gets pushed into the cross, which will be visible in the AHE signal (as demonstrated in Supplementary Note 3). 

The pinning of the DW at the entrance of the Hall cross happens due to the fact that the DW length has to increase in order to pass through the cross \cite{Ravelosona2005}. One way to overcome this problem, while still being able to use the legs for the AHE measurement, is to magnetically `cut-off' the legs using a technique called magnetic etching \cite{Franken2012}. In this technique a magnetic sample with perpendicular magnetic anisotropy (PMA) is exposed to Ga$^{+}$ ion irradiation with a relatively high dose ($40$ $\mu$C/cm$^{2}$ in this work), which is enough to destroy the magnetic anisotropy, but not enough to physically remove a significant amount of material. Using this technique the PMA in (part of) the legs is destroyed, causing it to become in-plane magnetized (or even paramagnetic), while the legs keep their conductive properties needed for the AHE measurement. In this way the AHE measurement can still be performed to measure the magnetization in the Hall cross area, while the legs become invisible for the DW. As a result, the DW will not be pinned at the entrance of the Hall cross since it does no longer need to increase in length when passing through the cross. 

\begin{figure}
	\includegraphics[scale=0.5]{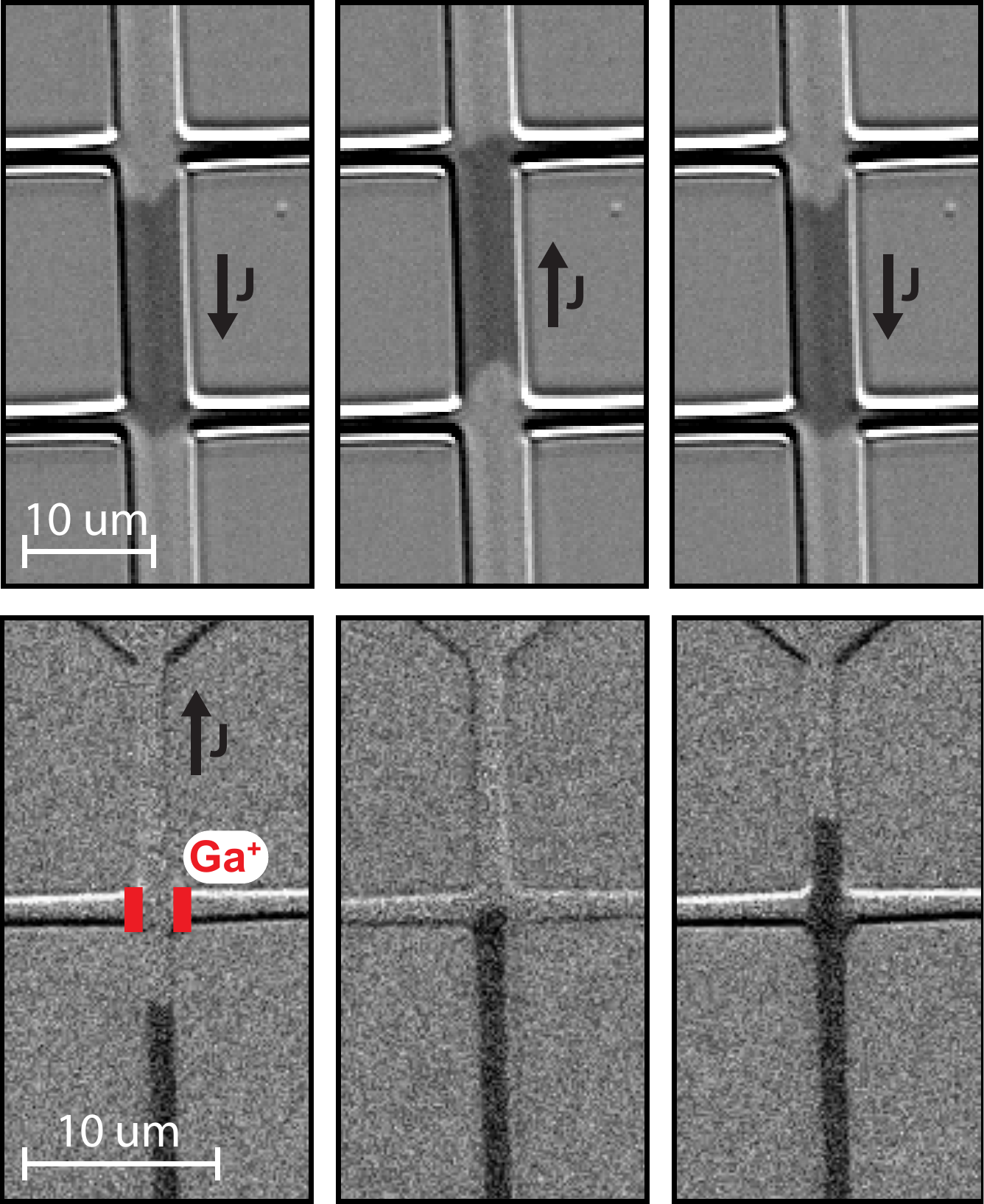}
	\caption{\textbf{DW pinning at a Hall cross and the effect of Ga$^{+}$ ion irradiation.} Top row: A Pt/Co/Gd wire with a down domain (dark) in an otherwise up (light) magnetized wire that is located between two Hall crosses. The three figures show the domain after three current pulses of alternating direction (direction indicated in figures). Bottom row: A Hall cross on a $2$ $\mu$m Pt/Co/Gd wire with a DW initially located below the cross. The red squares in the left figure represent the regions that are exposed to the Ga$^{+}$ ions. The figures show three snapshots of the DW moving through the wire by the SHE using a DC current (direction indicated in left figure).}
	\label{Fig:GaIrradiation}
\end{figure}

The effect of the Ga$^{+}$ ion irradiation on the DW propagation through the Hall cross is verified in the bottom row of Fig.\ \ref{Fig:GaIrradiation}. In these figures a Hall cross on a $2$ $\mu$m Pt/Co/Gd wire is shown, where the red squares in the left figure represent the regions that are exposed to the Ga$^{+}$ ions. Starting with a DW below the Hall cross in the left figure, three snapshots are shown of the DW moving through the wire by the SHE using a DC current. It is seen that with the legs being magnetically cut-off, the DW indeed is able to move past the Hall cross without getting pinned. 

\subsection*{\large Supplementary Note 3: On-the-fly AOS with different laser spot to (non-irradiated) Hall cross alignments}

A proof-of-concept measurement was presented in Fig.\ 2 of the main Letter, demonstrating on-the-fly single-pulse AOS and simultaneous SHE driven motion of magnetic domains in a single racetrack. In that measurement, the laser spot was aligned to the right side of the first Hall cross to prevent pinning of the DW's (see Fig.\ 2(a) of the main Letter), while a small overlap between laser pulse and the first Hall cross was maintained in order to verify if and when a domain was written. In this section, two similar measurement are presented; (i) with the laser spot centered at the center of the first Hall cross, clearly demonstrating the pinning of the DW at the entrance of the non-irradiated Hall cross, and (ii) with the laser spot aligned completely in between the two Hall crosses.

\begin{figure*}
	\includegraphics[scale=0.33]{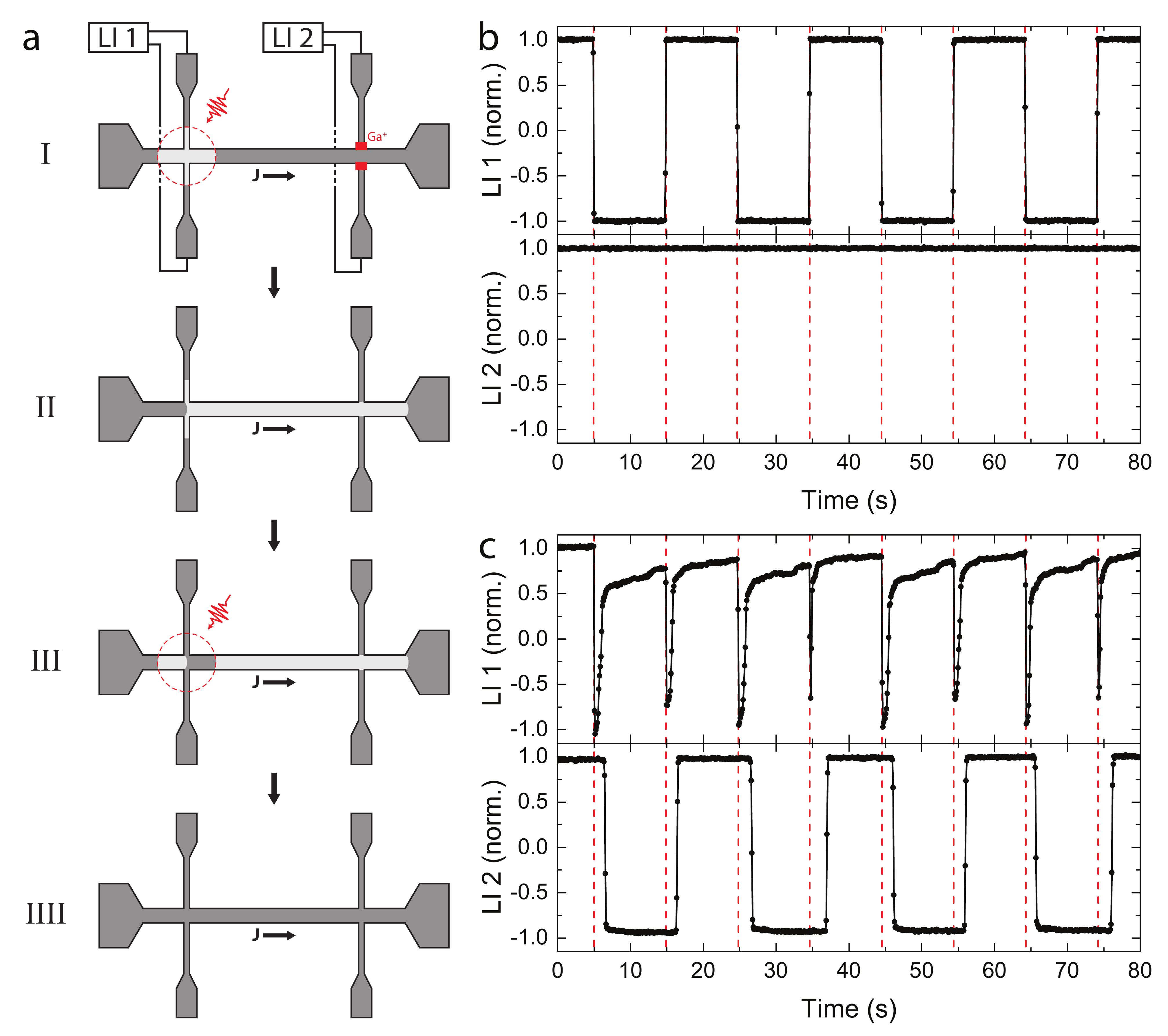}
	\caption{\textbf{On-the-fly AOS with the laser spot centered on the non-irradiated Hall cross.} (a) Illustration of the on-the-fly AOS measurement performed with the laser spot aligned to the center of the first Hall cross. The AHE signal in the Hall crosses is measured using lock-in amplifiers LI $1$ and LI $2$. The red dotted circle illustrates the region exposed by the laser pulse, and the red blocks indicate the regions exposed to Ga$^{+}$ ion irradiation. Figures I to IIII show different snapshots of the magnetization in the wire during a two-pulse cycle. (b,c) Measurement of the normalized AHE signal as a function of time in the first (top) and second (bottom) Hall cross while at the same time the first Hall cross is exposed to a train of linearly polarized laser pulses ($\approx 100$ fs) at a laser-pulse repetition rate of 0.1 Hz. The measurement presented in (b) was performed without any current sent through the wire, while a current of $+5.5$ mA was sent through the wire during the measurement presented in (c).}
	\label{Fig:ProofOfPrincipleOnCross}
\end{figure*}

An illustration of the on-the-fly AOS measurement performed with the laser spot aligned to the center of the first Hall cross is shown in Fig.\ \ref{Fig:ProofOfPrincipleOnCross}(a). The expected magnetic behavior is as follows; (I) With the magnetization in the wire initially saturated, the first (left) cross is exposed to a single laser pulse (red dotted circle), writing a magnetic domain that is larger than the Hall cross, thus creating a DW at either side of the cross. (II) Due to the SHE originating from the DC current that is continuously sent through the wire, both DW's (and thereby the domain) move along the current direction towards the second Hall cross. The DW written in between the two Hall crosses will reach the second cross, where it can pass the cross since its legs are magnetically cut-off using Ga$^{+}$ ion irradiation. The legs of the first Hall cross, however, have not been irradiated. Therefore, the DW written to the left of the first cross will get pinned at the entrance of the cross. The center of the DW is pushed into the cross by the SHE, and will be measurable by the AHE. (III) The next laser pulse will toggle the magnetization in the exposed area, creating two new DW's and reversing the polarity of the DW located in the first Hall cross. (IIII) The newly created DW's are moved along the wire by the SHE and annihilate with the previously written DW's, leaving the wire in the saturated state. This process will then repeat itself at the next laser pulse. 

First, a test measurement is performed without a DC current being sent through the wire, and using a repetition rate of 0.1 Hz to clearly see the effect of the single laser pulses. The result is presented in Fig.\ \ref{Fig:ProofOfPrincipleOnCross}(b), showing the normalized AHE signal of both crosses as a function of time. The top graph shows the AHE signal of the first cross, demonstrating clear and full single-pulse AOS of the magnetization in this cross, similar as shown in Fig.\ 1(b) of the main Letter. The bottom graph shows the AHE signal of the second Hall cross. As can be seen by the constant AHE signal at the initial saturation value, there is no effect of the laser pulses on the magnetization in the second cross, which is expected since there is no current sent through the wire, and thus no DW motion. 

Figure \ref{Fig:ProofOfPrincipleOnCross}(c) shows the result of a measurement with a DC current of $+5.5$ mA sent through the wire. Looking at the AHE signal of the first Hall cross (top graph), the signal looks much different than the signal measured without a DC current (Fig.\ \ref{Fig:ProofOfPrincipleOnCross}(b)). Before explaining the observed behavior in more detail, it is noted that the times at which the magnetic domains are written at the first Hall cross are still clearly visible by the sudden steps in the AHE signal (red dotted lines). Looking at the AHE signal in the second Hall cross (bottom graph), it can be seen that shortly after the domains are written, the magnetization in the second cross switches its direction, toggling up and down after each subsequent laser pulse. The toggling behavior corresponds to the DW that is written to the right of the first Hall cross that is transported along the wire by the SHE and passes through the second Hall cross. As shown in Fig.\ \ref{Fig:ProofOfPrincipleOnCross}(a), this DW alternates between an up-down and down-up DW on subsequent laser pulses, leaving the magnetization in the cross in the up and down state after the DW has passed, respectively. 

Coming back to the AHE signal measured in the first Hall cross (top graph Fig.\ \ref{Fig:ProofOfPrincipleOnCross}(c)), there are two observations that can be made. Firstly, the step in the AHE signal varies in size and has the same sign for every laser pulse. Secondly, the AHE signal after every even number of pulses is not equal to the initial saturation value of $+1$. Both observations will be discussed in the following.  

\begin{figure*}
	\includegraphics[scale=0.5]{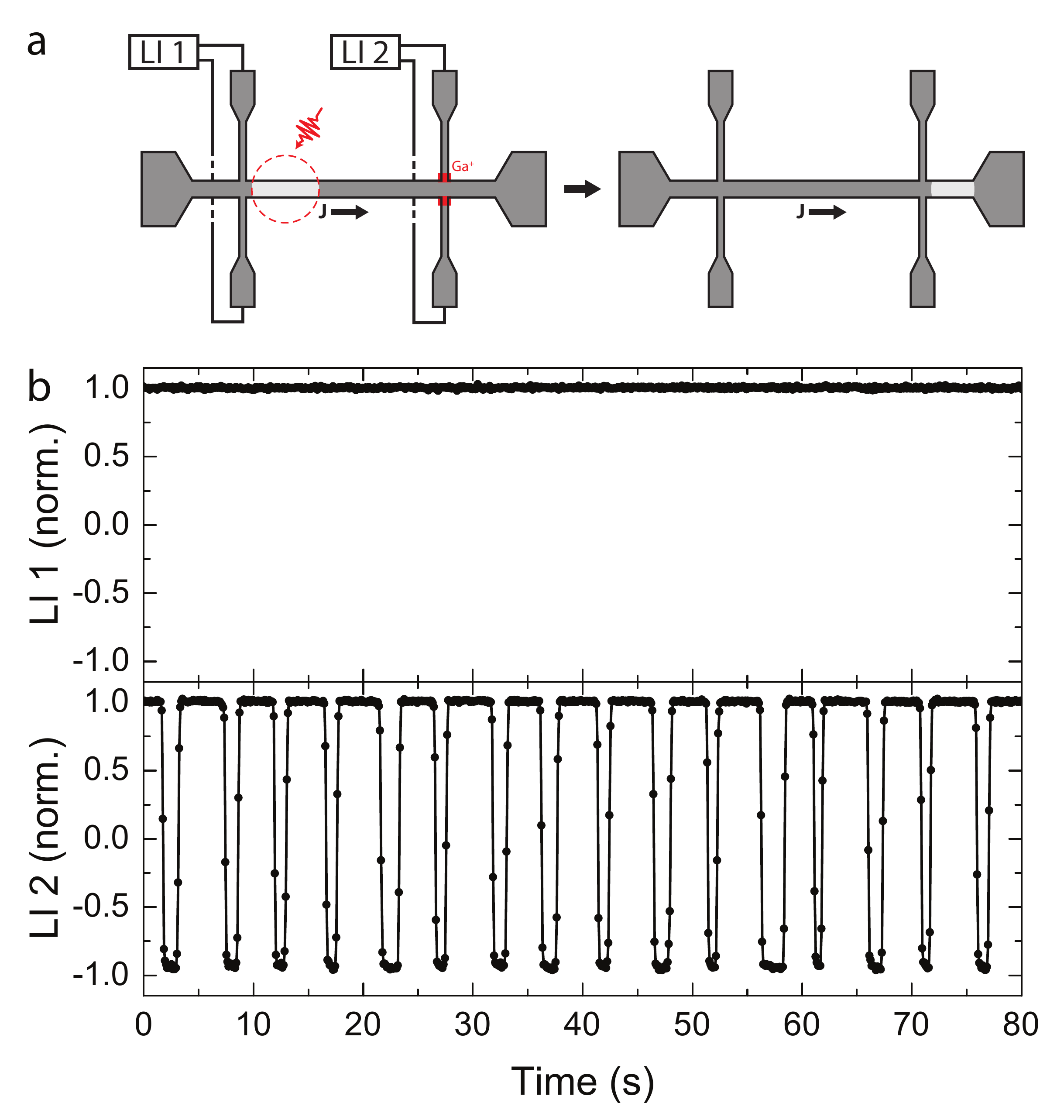}
	\caption{\textbf{On-the-fly AOS with the laser spot aligned in between the two Hall crosses.} (a) Illustration of the on-the-fly AOS measurement performed with the laser spot aligned in between the two Hall crosses of the Pt/Co/Gd wire. The AHE signal in the Hall crosses is measured using lock-in amplifiers LI $1$ and LI $2$. The red dotted circle illustrates the region exposed by the laser pulse, and the red blocks indicate the regions exposed to Ga$^{+}$ ion irradiation. (b) Measurement of the (normalized) AHE signal as a function of time in the first (top) and second (bottom) Hall cross while at the same time the wire is exposed to a train of linearly polarized laser pulses ($\approx 100$ fs) at a laser-pulse repetition rate of 0.2 Hz and a DC current of $+5.5$ mA is sent through the wire.}
	\label{Fig:ProofOfPrincipleNoOverlap}
\end{figure*}

It can be seen that the very first laser pulse switches the magnetization from the initial saturated up ($+1$) direction to the down ($-1$) direction. In contrast to the case without the DC current, the AHE signal immediately rises again towards $\approx +0.5$. This corresponds to the DW written to the left of the first Hall cross being transported along the wire by the SHE and getting pinned at the entrance of the cross (Fig.\ \ref{Fig:ProofOfPrincipleOnCross}(a).II). The center of the DW is being pushed inside the cross by the SHE, causing about $75\%$ of the Hall cross area being switched back to the initial saturation direction. After this initial (fast) response, the center of the DW keeps getting pushed further inside the cross at a slower rate, causing a further (slow) increase of the AHE signal. When the second pulse hits the sample, most of the Hall cross area ($\approx 85\%$) is switched back to the initial saturation direction (Fig.\ \ref{Fig:ProofOfPrincipleOnCross}(a).III). Therefore, the second step in the AHE signal has the same sign as for the first pulse, however, the size of the step is smaller since it only corresponds to the net switched magnetization. This measurement clearly demonstrates the pinning of the DW at the entrance of the Hall cross when the legs are not magnetically cut-off using Ga$^{+}$ ion irradiation.

Taking a closer look at the AHE signal in the first Hall cross just before the third pulse arrives, it can be seen that the signal is not back to the saturation value of $+1$, but is close to $+0.9$ which corresponds to $95\%$ of the first Hall cross having its magnetization up. Looking back at Fig.\ \ref{Fig:ProofOfPrincipleOnCross}(a), it was expected that the DW written on the left side of the cross by the second pulse (or every even amount of pulses) would annihilate with the DW pinned in the cross (written by the previous pulse), leaving the Hall cross in the initial saturated state. This discrepancy is believed to be the result of the center of the DW that is being pushed into the cross not reaching a steady position before the next laser pulse arrives. This could result in the formation of thin rings with alternating magnetization direction in the first Hall cross that slowly move along with the current. The presence of these rings cause the AHE signal not to reach the saturation value of $+1$ after every even amount of pulses. 

Lastly, the same on-the-fly AOS measurement is performed, but now with the laser spot aligned completely in between the two Hall crosses, as shown in Fig.\ \ref{Fig:ProofOfPrincipleNoOverlap}(a), and a laser-pulse repetition rate of 0.2 Hz. The AHE signal of both Hall crosses as a function of time is shown in Fig.\ \ref{Fig:ProofOfPrincipleNoOverlap}(b). The magnetic domains are written to the right side of the first Hall cross, and they are transported along the current direction. This means that none of the domains or DW's reach the first Hall cross, resulting in the observed constant saturation value of the AHE signal in the first Hall cross (top graph). Looking at the AHE signal of the second Hall cross (bottom graph), it can be seen that all the optically written domains pass the second cross, as was also demonstrated in the measurement presented in Fig.\ 2(b) of the main Letter. Also seen in both measurements is the variation in the width of the domains when they pass the second cross (proportional to the time between down and up switch in the AHE signal). This shows that the variation in domain width measured in Fig.\ 2(b) of the main Letter is not dominated by the slight overlap between laser spot and first Hall cross, but is more likely to be the result of random pinning along the wire, as was already stated in the main Letter.

\end{document}